\documentstyle[preprint,epsf,aps]{revtex}

\begin{document}

\tightenlines

\draft

\title{Burgers Turbulence and the Continuous Spontaneous 
Localization Model}

\author{L. F. Santos $^1$ and C. O. Escobar $^2$}

\address{$^1$ Departamento de F\'{\i}sica Nuclear \\
Instituto de F\'{\i}sica da Universidade de S\~ao Paulo, 
C.P. 66318, cep 05389-970 \\
S\~ao Paulo, S\~ao Paulo, Brazil\\
lsantos@charme.if.usp.br \\
$^2$ Departamento de  Raios C\'osmicos e Cronologia \\
Instituto de F\'{\i}sica Gleb Wataghin \\
Universidade Estadual de Campinas, C.P. 6165,  cep 13083-970\\
Campinas, S\~ao Paulo, Brazil\\
escobar@ifi.unicamp.br}

\maketitle

\begin{abstract}

There is a striking convergence between Burgers turbulence and the 
continuous spontaneous localization [CSL] model
of quantum mechanics. In this paper, we exploit this analogy 
showing the similarities in the physics of these two 
apparently unrelated problems. It is hoped that the kind 
of analogy we introduce
here  may lead to important developments in both areas.\\ 
\end{abstract}


\section{Introduction}

The study of nonlinear growth processes has recently become of 
central interest in physics, an
example is provided by the KPZ equation 
\cite{Forster,KPZ}. Originally devised
to describe crystal growth, it has been since then applied to a wide range
of systems, from bacterial growth \cite{Schnerb,Vicsek} to directed
polymers \cite{Mezard}. One of the essential features of the KPZ equation is
its nonlinear term introduced in order to account for lateral growth beyond
the linear approximation such as described by the Edwards-Wilkinson model
\cite{Zhang}.

The KPZ equation is given by

\begin{equation}
\frac{\partial h({\bf x},t)}{\partial t} = \nu \nabla ^2 h({\bf x},t)
+ \frac{1}{2} (\nabla h({\bf x},t))^2 +\phi ({\bf x},t),
\end{equation}
where $h({\bf x} ,t)$ is the surface position, $\nu $ is the viscosity and
$\phi ({\bf x} ,t)$ is a space and time dependent noise.

There is an interesting connection between the KPZ equation and the Burgers
equation. The latter is a nonlinear diffusion equation for the velocity
field
of a fluid in $N$ dimensions. The velocity is a gradient field,
${\bf v} = - {\bf \nabla }h$, and its equation is written as

\begin{equation}
\frac{\partial {\bf v}}{\partial t} + ({\bf v} .{\bf \nabla }) {\bf v} =\nu
\nabla ^2 {\bf v} + {\bf f}({\bf x},t) ,
\end{equation}
where the external stirring force is given by ${\bf f} = - {\bf
\nabla }\phi $ and its correlation is \cite{Bouchaud}

\begin{equation}
\langle f^{\mu } ({\bf x}, t) f^{\nu } ({\bf x}, t') \rangle  = 
\epsilon \delta (t-t') \left[ \delta ^{\mu \nu} - 
\frac{({\bf x} - {\bf x'})^{\mu }
({\bf x} - {\bf x'})^{\nu }}{N \Delta ^2 } \right] 
\exp \left[ - \frac{({\bf x} - {\bf x'})^2 }{2N \Delta ^2 } \right].
\end{equation}
$\epsilon $ is the energy injected into the fluid per unit time and unit
mass and $\Delta $ is the length scale at which energy is injected.

In terms of the random potential the correlation function is

\begin{equation}
\langle \phi ({\bf x}, t) \phi ({\bf x'}, t') \rangle = 
\epsilon \Delta ^2 N \delta (t-t') 
\exp \left[-\frac{({\bf x} - {\bf x'})^2 }{2N\Delta ^2 }\right] .
\label{phi}
\end{equation}

Through the Hopf-Cole
transformation \cite{Mezard},
${\bf v} = -2\nu [{\bf \nabla }Z({\bf x}, t)]/Z({\bf x}, t)$,
the Burgers equation can be put into a linear 
form with a multiplicative noise

\begin{equation}
\frac{\partial Z({\bf x},t)}{\partial t} = \nu \nabla ^2 Z({\bf x},t) +
\frac{\phi ({\bf x}, t)}{2\nu } Z({\bf x},t).
\label{z}
\end {equation}

The above equation is a Schr\"odinger equation in the imaginary time. A
similar equation for real time has been considered by several authors in a
different context, such as the description of quantum open systems
\cite{Diosi,Gisin} or the spontaneous collapse of the wave function [CSL
model] in an attempt to solve the quantum measurement problem
\cite{GPR,Pearle}.

In a previous article \cite{PRA}, we have presented an analogy between the
CSL model and enhanced diffusion.  We started from a comparison of our
result for $\langle x^2 \rangle $ in the beable interpretation of the CSL
model with similar results obtained by
Shlesinger {\it et al} \cite{Shlesinger}. In the present paper,
we show that this analogy extends much further by exploring a similarity
between the CSL
modified Schr\"odinger equation and the KPZ, Burgers equations.

The first section establishes an analogy between the KPZ and
the CSL models, which provides a dictionary we use throughout the
paper to move from one system to another and which allows us to introduce a
Reynolds number for the CSL model. In section two we develop the stochastic
picture underlying our beable interpretation of the CSL model and connect it
with the Burgers equation through the velocity field. This enables us to
introduce the intermittency corrections and  a Fokker-Planck equation [FPE]
corresponding to the stochastic equation for the CSL model in the random
force approximation \cite{Mezard}. An analysis of this system of equations
allows us to introduce a time scale characterizing the dominance of
enhanced diffusion over the standard (Wiener) diffusion. 
Some of the results are
numerically illustrated with the sample paths of our stochastic differential
equation.
Section III presents our conclusions.

\section{A Dictionary for KPZ, Burgers and CSL}

The CSL model modifies the Schr\"odinger  equation by introducing a
multiplicative noise in the evolution of the wave function.
In one dimension, the evolution equation in the Stratonovich form for a free
particle of mass $M$ is given by

\begin{equation}
\frac{\partial \psi (x,t)}{\partial t}= 
\left\{ i \nu \frac{\partial ^2}{\partial x^2} -
\lambda + \sqrt{\gamma } \int dz w(z,t)G(x-z) \right\}
\psi (x,t),
\label{csl}
\end{equation}
where $w(z,t)$ is a white noise, so that  $\langle
w(z,t) \rangle = 0$ and $\langle w(z,t)w(z',t')\rangle =\delta (z-z') 
\delta (t-t')$ and

\begin{equation}
G(x-z) = \sqrt{\frac{\alpha }{2\pi }} \exp \left[ -\alpha
\frac{(x-z)^2 }{2}\right]
\end{equation}
characterizes the localization of the wave function. The diffusion
constant $\nu $ is equal to $\hbar /2M$ and $\gamma$ is related to the
localization scale $1/\sqrt{\alpha } $ and the frequency of collapse
$\lambda $ as $\gamma  = \lambda \sqrt{4\pi /\alpha }$. These parameters are
chosen in such a way that the new evolution equation does not give different
results from the usual Schr\"odinger unitary evolution for microscopic
systems with few degrees of freedom, but when a macroscopic system is
described there is a fast decay of the macroscopic linear superpositions
which are quickly transformed into statistical mixtures \cite{GPR,GRW}.

The above modified Schr\"odinger equation is similar to eq.(\ref{z}) in
the imaginary time. From this analogy the noise term may be written as
$\phi (x,t) = 2\nu \sqrt{\gamma }\int dz w(z,t)G(x-z) $, which gives the
following correlation function

\begin{equation}
<\phi (x,t) \phi (x',t')> = 4 \nu ^2 \lambda \delta (t-t') 
\exp [-\alpha \frac{(x-x')^2 }{4}].
\end{equation}
Comparing this equation with eq.(\ref{phi}), we identify the injection
length scale $\Delta $ with $\sqrt{2/\alpha }$ and the injected energy
$\epsilon $ with $2\nu ^2 \alpha \lambda $ and can then easily obtain the
Reynolds number for the CSL model by using the relation \cite{Mezard}: 
$Re =(\epsilon \Delta ^4 /\nu ^3 ) ^{1/3}$, which gives
$Re = (8\lambda /\alpha \nu )^{1/3}$.
For large Reynolds number we are in the domain of fully developed
turbulence, which in the CSL case corresponds to a quantum system undergoing
frequent collapses as $\lambda $
is large.

The analogy between turbulence and the CSL model also provides a
clarification regarding the issue of energy
non-conservation
in the original model for the collapse of the wave function 
proposed by Ghirardi, Rimini
and Weber [GRW] \cite{GRW}. The celebrated Kolmogorov analysis of
1941 [K41] \cite{Kolmo}, identifies $\Delta $ as the scale at which energy
$\epsilon $ is injected into the system and a
much smaller scale $l_{d} $ at which energy is dissipated. For distances $r$
such that $l_{d}
\ll r\ll \Delta $, we have what is called the inertial regime, where energy
is transferred to ever smaller length scales.
The analogous result for the collapse model has that the energy is injected
at the localization
scale $1/\sqrt{\alpha } $, which in turn is much larger than the atomic
scale.
In the framework of the collapse model, energy non-conservation coming from
the collapse has always been discussed in terms of the magnitude of the
effect
itself and the constants were chosen in such a way as to make it a hardly
observable effect at all. Here we stress, in analogy with turbulence, that
the system does not gain energy indefinitely, for the energy injected in the
collapse will be dissipated at the atomic scale. The disparity of scales is
such that this fact was naturally unnoticed by the proponents of the
collapse model.

For completeness, let us end this section adding that it is in the inertial
regime that Kolmogorov's scaling results

\begin{equation}
\langle |{\bf v}({\bf x} - {\bf r}, t) - 
{\bf v}({\bf x}, t)|^p \rangle 
\sim r^{\frac{p}{3}} ,
\end{equation}
though it fails when $p$ is larger than 3, a phenomenon known as
intermittency, to which we shall return in the next section.

\section{Developing the analogy}

We have recently analyzed the CSL model from a microscopic point of view
\cite{PRA}.
In order to do so we used Vink's treatment \cite{Vink}, which shows that two
alternative interpretations of quantum mechanics that treat position as a
classical concept, the
causal interpretation due to Bohm \cite{Bohm}
and the stochastic
interpretation due to Nelson \cite{Nelson}, are actually
particular cases of Bell's 'beable'
interpretation \cite{Bell}. The
beable interpretation is an attempt by Bell to treat physical quantities
that
exist independently of observation and therefore can be
assigned well defined values. His approach used fermion
number, a discrete quantity, and Vink extended it to
any observable that takes discrete values on small scales. One starts from
the equation for the probability density $P_{m} (t)$ on a given basis

\begin{equation}
\partial _{t} P_{m} = \sum_{n} J_{mn},
\label{disc}
\end{equation}
where the source matrix $J_{mn}$
is given by

\begin{equation}
J_{mn} (t) = 2 Im \{\langle \psi (t)| O_{m} \rangle 
\langle O_{m} |H| O_{n} \rangle  
\langle O_{n} | \psi (t)\rangle  \}.
\end{equation}

From a stochastic point of view, the  probability distribution
of $O_{m} $ values satisfies the master equation

\begin{equation}
\partial _{t} P_{m} = \sum_{n} ( T_{mn} P_{n} - T_{nm} P_{m} ),
\label{me}
\end{equation}
where $T_{mn} dt$ is the transition probability for jumps from state $n$ to
state $m$.

To reconcile the quantum and stochastic views we equate (\ref{disc}) and
(\ref{me}):

\begin{equation}
J_{mn} = ( T_{mn} P_{n} - T_{nm} P_{m} ),
\label{j}
\end{equation}
with $T_{mn} \geq 0$ and  $J_{mn} = - J_{nm} $.

There is great freedom to find solutions of
eq.(\ref{j}). Bell chooses a particular one
for $n \neq m$,

\[ T_{mn} = \left\{
 \begin{array}{ll}
  J_{mn} /P_{n} &    J_{mn} > 0 \\
  0 &  J_{mn} \leq 0 .
 \end{array}
\right. \]

Restricting the position of a particle in one
dimension to the sites of a lattice, $x=an$, with $n = 1,...,N$ and
$a$ the lattice distance, it follows from the discrete version
of the Schr\"odinger equation that $J_{mn} $ is given by

\begin{equation}
J_{mn} = \frac{1}{Ma } \{[S (an )]' P_{n} \delta _{n, m-1 }
- [S (an)]' P_{n}  \delta _{n, m+1 }\} ,
\end{equation}
where use was made of the polar form of the wave function 
$\psi = R e^{(iS/\hbar )} $ and
$\psi (x+a) $ is expanded up to first order in $a$. In the expression
above $[S(an)]' = [S(an+a) - S(an)]/a $.

For forward movement, Bell's choice becomes

\begin{equation}
T_{mn} = \frac{[S(an)]'}{Ma} \delta _{n, m-1} .
\end{equation}
This term leads to transitions only between neighboring sites.
However, we could also have added to $T_{mn}$ the solution of the
homogeneous equation derived from eq.(\ref{j}):

\begin{equation}
T^{o}_{mn} \propto
\exp \left\{ -\left[ m-n - \frac{2\sigma ln(P_{m} /P_{n} )}{4(m-n)}
\right] ^2 /2\sigma \right\} .
\end{equation}
This extra term introduces transitions between more distant sites.

Assuming $\sigma $ sufficiently small, we can approximate
$[\ln (P_{m} /P_{n})]/(m-n)$ by $2a[R(an)]'/R(an) $ and arrive at the
following
Langevin equation for the particle position in the continuum limit

\begin{equation}
\dot{x} = v(x,t) +
(\beta \sigma a^2 )^{\frac{1}{2} } \eta (t) ,
\label{nelson}
\end{equation}
where $\beta $ is a free parameter, $\eta (t)$ is a white noise, such that
$\langle \eta (t) \rangle = 0$, $\langle \eta (t) \eta (t') \rangle = 
\delta (t-t')$ and

\begin{equation}
v(x,t) =\left[ (\beta
\sigma a^2 ) \frac{1}{R(x,t)} \frac{\partial R(x,t)}{\partial x} +
\frac{1}{M} \frac{\partial S(x,t)}{\partial x} \right].
\end{equation}

Eq.(\ref{nelson}) coincides with Nelson's stochastic equation with
$\beta \sigma a^2 = 2\nu $. A similar equation was obtained in our analysis
of the microscopic dynamics of the
CSL model. This is so because the new terms in the source matrix $J_{mn}$
coming
from the modified Schr\"odinger equation (\ref{csl}) do not contribute to
the displacement $dx$ \cite{PRA}.

The connection between the CSL (beable) equation and the Burgers equation is
now straightforward. In imaginary time $S(x,t)=0$ and $v=2\nu (\nabla R)/R
$, which, replacing $R$ by $Z$, corresponds to the Hopf-Cole transformation.
The velocity satisfies then the Burgers equation if we assume the relation
between the random force and the random potential without the minus sign
$f=\nabla \phi $ (a point that was already noticed by Garbaczewski {\it et
al} \cite{Garba}).

To simplify the following calculations, the Hamiltonian is set equal to zero
as done in \cite{tail,Squires}. The solution of the modified
Schr\"odinger equation is easily obtained

\begin{equation}
\psi(x,t) = \exp (-\lambda t) \exp \left[ \sqrt{\gamma } \int _{0}^{t} dt'
\int dz G(x-z) w(z,t') \right] .
\end{equation}

By choosing the initial wave function as a single Gaussian

\begin{eqnarray}
\psi (x,0) &=& \frac{1}{(2 \pi \Delta x ) ^{\frac{1}{4}} } 
\exp \left[ -\frac{(x-\langle x\rangle )^2 }{4 \Delta x } \right] \\
           & & \exp \left\{ \frac{i}{\hbar }
\left[ \frac{(\frac{x^2 }{2} - x\langle x\rangle )
\sqrt{\Delta x \Delta p -\hbar ^2 /4}}
{\Delta x} + \langle p\rangle x \right] \right\} , \nonumber
\end{eqnarray}
where $\Delta x = \langle x^2 \rangle - \langle x \rangle ^2 $ 
and $\langle x\rangle $ is the mean 
value of $x$ (the same is valid for $p$), 
the stochastic differential equation for position is

\begin{equation}
\dot{x} = D_{S} 
+ 2\nu \sqrt{\gamma } 
\left\{ \int _{0}^{t} dt' \int dz w(z,t') 
[-\alpha (x-z)]G(x-z) \right\}  + 
\sqrt{2\nu } \eta (t) .
\label{dx}
\end{equation}
This equation describes the evolution of the position of a tracer 
in a turbulent medium.
The first term on the right hand side 
describes a single free particle
deterministic evolution and we choose $D_{S} $ as 
a short notation for the term derived from the initial 
wave function 
$D_{S} = [2\nu \nabla R_{S} (x,0)/R_{S} (x,0) + \nabla S_{S} (x,0) /M]
= [\langle x\rangle -x][\nu /\Delta x - \sqrt{\Delta x \Delta p - 
\hbar ^2 /4} /M \Delta x] + \langle p\rangle /M$. 
The two other terms describe the stochastic processes. 
The last term corresponds to a Brownian diffusion and the 
second one is responsible for the $t^3 $ behavior of the mean 
square displacement, which is the same time dependence obtained 
by Richardson in his pioneering studies of turbulence \cite{Richardson}. 
Notice that the coefficient of $\langle x^2 \rangle $ corresponds to 
the injected energy $2 \alpha \lambda \nu ^2 $ as expected from 
hydro dynamical turbulence \cite{Procaccia} and coincides with the 
result in the previous section.

Moreover, the time dependence and nonlocal character of the second 
term on the r.h.s. of the equation above are in accordance with the
 concept of L\'evy walk as introduced by Shlesinger {\it et al} 
\cite{Shlesinger} when studying the phenomenon of enhanced diffusion.
The basic difference between the more familiar L\'evy flight \cite{Levy} 
and  L\'evy walk is 
that for the latter, although the walker visits all sites visited 
by the flight, the jumps do not occur instantaneously,
 but there may be a time delay before the next jump. By introducing time,
Shlesinger {\it et al}
obtained an integral transport equation involving a
scaled memory which is nonlocal in space and time.
Contrary to the infinite mean square displacement
obtained in a L\'evy flight, the solution of
such transport equation leads to a finite mean square displacement
such as the one obtained by Richardson.

From eq.(\ref{dx}), it follows that momentum satisfies

\begin{equation}
\dot{p} =2 M\nu \sqrt{\gamma }
\int dz w(z,t)
[-\alpha (x-z)] G(x-z) .
\label{dp}
\end{equation}
The stochastic process for momentum derives from the introduction of the
random potential in the modified Schr\"odinger equation, which in turn is
responsible for the localization of the wave function. This process vanishes
when the GRW parameters go to zero.

The FPE corresponding to eq.(\ref{dx}), which is a stochastic differential
equation with a colored noise introduced by $v(t)$, is obtained as done by
Zoller {\it et al} \cite{Zoller,Risken}. However, as we are in the inertial
regime we use the random force approximation \cite{Larkin,Mezard}: $\phi
\sim A(t) + f(t) x(t)$ getting the following simplified equation for
momentum

\begin{equation}
\dot{p} = \hbar \sqrt{\frac{\alpha \lambda }{2}} \eta (t) .
\label{dps}
\end{equation}

The FPE is then

\begin{eqnarray}
\frac{\partial P (x,p,t)}{\partial t} &=&
- \frac{\langle p\rangle}{M}  \frac{\partial P(x,p,t) }{\partial x} 
\nonumber \\
                                      &+&
\left[ 
\frac{\hbar }{2M} \frac{\partial ^{2} }{\partial x^2 } 
+ \sqrt{\frac{\hbar ^3 \alpha \lambda}{2M}} 
\frac{\partial ^2 }{\partial x \partial p} + 
\frac{\hbar ^2 \alpha \lambda}{4} \frac{\partial ^2 }{\partial p^2 } 
\right] P(x,p,t) .
\label{fpe}
\end{eqnarray}

A nice feature of our model is to obtain the above phase-space equation 
of evolution. Differently from Richardson \cite{Shlesinger}, we started 
from purely theoretical arguments and took into account the 
discontinuous nature of the particle velocity. Although eq.(\ref{fpe}) is
the same one as obtained by GRW \cite{GRW}, we believe it is an
oversimplification of the problem and the correct FPE to be used when
studying the diffusion of a tracer should come from the complete equation
for momentum, eq.(\ref{fpe}). This is so because the random force
approximation is deemed to be incorrect \cite{Mezard}, since within it there
is no intermittency. However, as we show below, there is a way out of this
shortcoming which allows us to obtain the intermittency corrections even in
the case of the random force approximation.

\subsection{Intermittency corrections}

Having eqs. (\ref{dx}) and (\ref{dp}) we can now proceed to obtain the
Mandelbrot intermittency corrections \cite{Mandelbrot} to Richardson's law.
In order to
do so we replace a white noise in time by
an affine one \cite{Mandelbrot2} called fractional Brownian noise

\begin{equation}
\langle w (z,t)\, w (z',t') \rangle = t^{A-1} \delta (t-t') 
\, \delta (z-z'),
\label{ww}
\end{equation}
which gives for the anomalous diffusion
term \cite{PRA}

\begin{equation}
\langle x^2 (t)\rangle \sim  t^{A-1+3} .
\label{xno}
\end{equation}
This corresponds to one of the intermittency
corrections obtained by Shlesinger {\it et al}  \cite{Shlesinger}
provided we identify $A-1$ with
$3 \mu /(4- \mu ) $, where $\mu = E - df $, $E$ being
the Euclidean dimension and $df$, the fractal dimension.

For the momentum variable the non-white noise gives

\begin{equation}
\langle p^2 \rangle \sim t^A ,
\label{pno}
\end{equation}
which leads to the scaling relation obtained by Shlesinger {\it et al}
\cite{Shlesinger} for the root-mean-square velocity.

We now see that, within the random force approximation, the intermittency
corrections can still be obtained as eq.(\ref{ww})leads to $\langle \eta
(t)\, \eta (t') \rangle = t^{A-1} \delta (t-t') $.

\subsection{Time scales}

Neglecting the deterministic term in eq.(\ref{dx}), we find the mean square
displacement $\langle x^2 \rangle $

\begin{equation}
\langle x^2 \rangle = 2\nu t + \frac{2}{3} \alpha \lambda \nu ^2 t^3 ,
\label{tenh}
\end{equation}
which has two contributions: the usual Brownian one, coming from the 
$\eta (t) $ term, and the enhanced diffusion \cite{PRA}, arising from the
multiplicative noise in the modified Schr\"odinger equation. In
fig.1 we illustrate, with sample paths from eq.(\ref{dx}), the enhanced
diffusion vis-\`a-vis the Brownian diffusion. 
We see that the numerical simulation exhibits the expected 
behavior of enhanced diffusion 

Comparison of the two terms in eq.(\ref{tenh}) allows 
us to determine the time scale beyond 
which the enhanced
diffusion dominates over the Brownian one

\begin{equation}
t_{enh} > \sqrt{\frac{3}{\alpha \lambda \nu }}
\end{equation}

Using the GRW parameters
\cite{GRW}: $\alpha = 10^{10} cm^{-2} $, $\lambda $(micro) $= 10^{-16}
s^{-1} $, $\lambda $(macro) $=10^{7} s^{-1} $, $M$(micro) $= 10^{-23} g$,
$M$(macro) $= 1 g$, we estimate the time scale, which is 
approximately $2.4x10^{5} s$, independently of the macro or 
microscopic nature of the system.

A second time scale is given by the characteristic collapse time, which is
of the order of $\lambda ^{-1} $. It does depend on the macro/microscopic
nature of the system. Fig.2 displays the two time scales. The dashed line
indicates the Schr\"odinger evolution, by which we mean the evolution of the
wave function undisturbed by the collapse term. For the microscopic system
enhanced diffusion manifests itself even before collapse occurs, opening an
interesting window for experimental tests of this scenario, which would help
to put stricter bounds on the parameters of the GRW/CSL models
\cite{exp}.

\section{Conclusions}

In this article we found further points of contact between turbulence and
the CSL model, identifying the CSL counterparts of important quantities in
turbulence, such as the Reynolds number and the injected energy.

A microscopic stochastic picture for the CSL model allowed us to show that
the velocity field of the tracer satisfies the Burgers equation.

Finally a study of the different time scales governing enhanced and Brownian
diffusion indicates that there is a region 
of parameter space of the CSL model
amenable to experimental test, an 
investigation we plan to follow in a forthcoming
publication.

\acknowledgments
The authors acknowledge the support of the 
Brazilian Research Council, CNPq.

\begin{figure}[ht]
\begin{center}
\epsffile{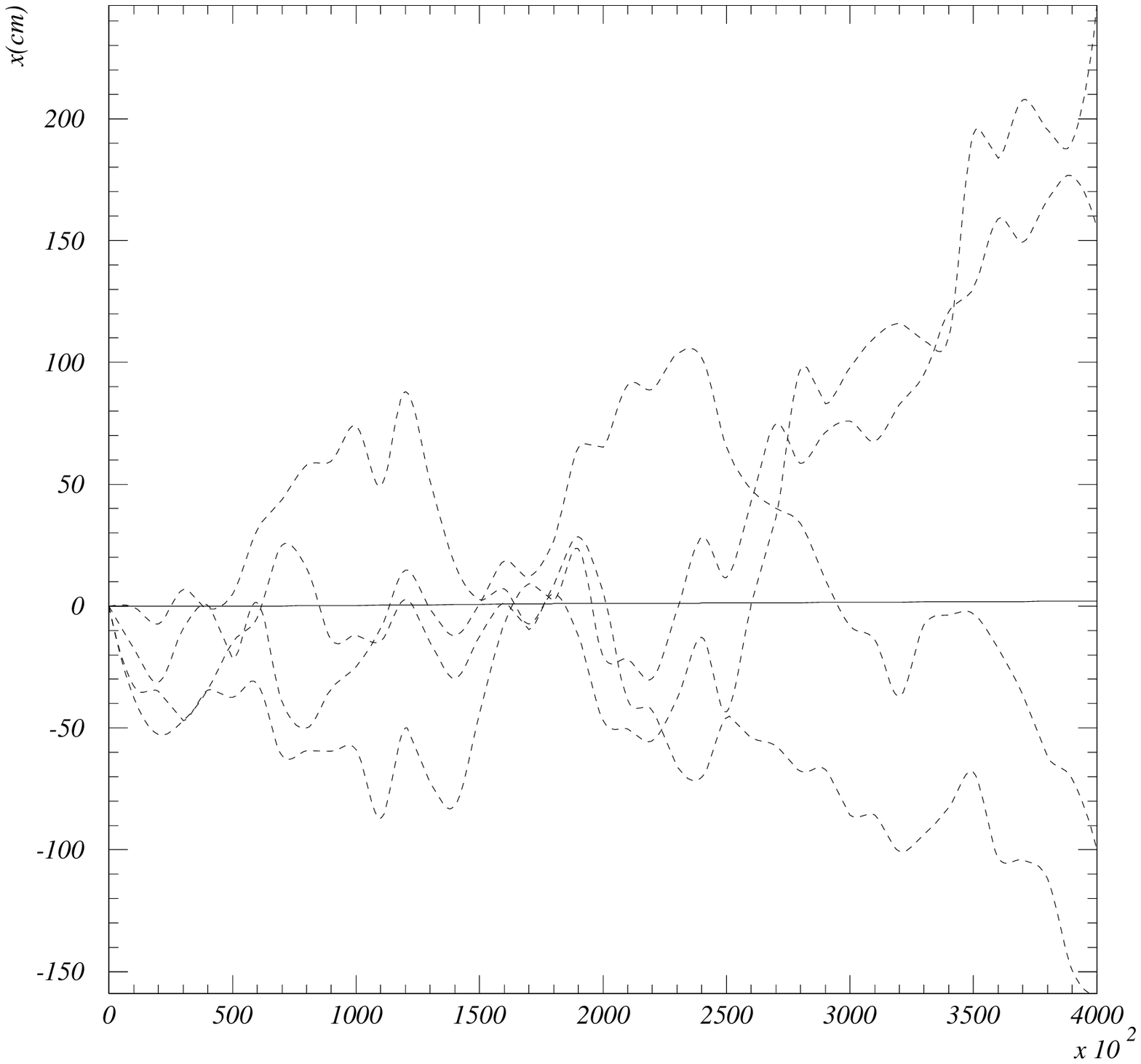}
\end{center}
\caption{The curves were obtained using the algorithm for the 
numerical integration of stochastic equations with colored noise 
developed by Fox [31]. The horizontal axis corresponds to time 
in seconds.
The full line represents the average value
of $x$, which was obtained after 5000 realizations, and the 
dashed lines correspond to 4 sample paths. The parameters are
those for a macroscopic particle 
and the time step is $\Delta t = 10^{4} s$.}
\label{fig1}
\end{figure}

\begin{figure}[ht]
\epsffile{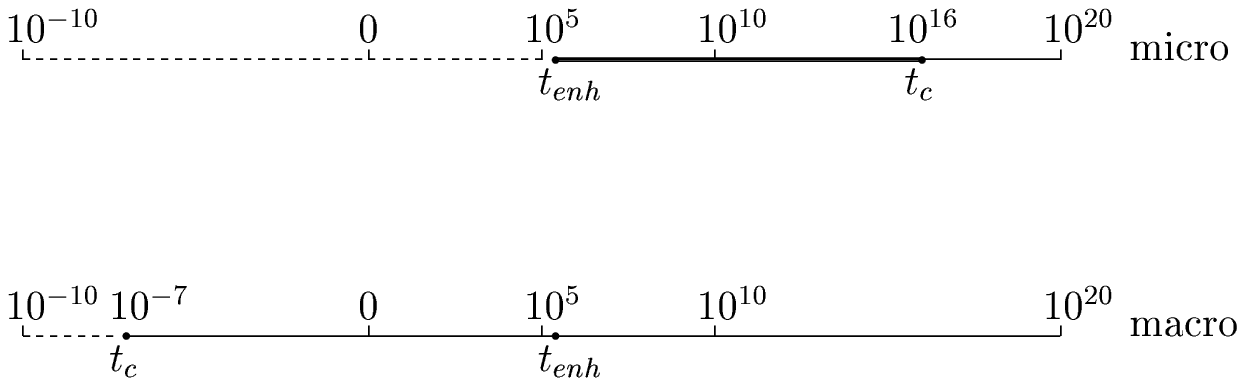}
\caption{The two lines give the time in seconds for 
the evolution of a microscopic system 
 and a macroscopic one.
The dashed lines correspond to the evolution of the wave function
undisturbed by the collapse term. The thick line, which only appears
for the microscopic system, corresponds to the time gap at which 
the CSL model may be tested.}
\label{fig2}
\end{figure}

\end{document}